\documentclass[aps,prl,showpacs,
  superscriptaddress,
  floatfix,
  twocolumn,
secnumarabic,amssymb,amsmath]{revtex4}

\usepackage{graphicx}
\usepackage{dcolumn}
\usepackage{amsmath}
\usepackage{amsfonts}
\usepackage{amssymb}
\usepackage{bm}
\usepackage{color}
\usepackage{ulem}
\usepackage{longtable}
\usepackage[colorlinks=true,linkcolor=blue,pdfstartview=FitH,pdfauthor=Gross]{hyperref}

\begin{document}

%\preprint{D. Venkateshvaran {\it et al.}, version: 05 April 2008}

\title{Anomalous Hall Effect in Magnetite: Universal Scaling Relation Between Hall and Longitudinal Conductivity in
Low-Conductivity Ferromagnets}

\author{Deepak Venkateshvaran}
 \affiliation{Walther-Mei{\ss}ner-Institut, Bayerische Akademie der
              Wissenschaften, 85748 Garching, Germany}
 \affiliation{Materials Science Research Centre, Indian Institute of Technology
                Madras, Chennai 600036, India}

\author{Wolfgang Kaiser}
 \affiliation{Walther-Mei{\ss}ner-Institut, Bayerische Akademie der
              Wissenschaften, 85748 Garching, Germany}

\author{Andrea Boger}
 \affiliation{Walther-Mei{\ss}ner-Institut, Bayerische Akademie der
              Wissenschaften, 85748 Garching, Germany}

 \author{Matthias Althammer}
 \affiliation{Walther-Mei{\ss}ner-Institut, Bayerische Akademie der
              Wissenschaften, 85748 Garching, Germany}

\author{M.S. Ramachandra Rao}
 \affiliation{Materials Science Research Centre, Indian Institute of Technology
                Madras, Chennai 600036, India}

\author{Sebastian~T.~B.~Goennenwein}
 \affiliation{Walther-Mei{\ss}ner-Institut, Bayerische Akademie der
              Wissenschaften, 85748 Garching, Germany}

\author{Matthias Opel}
\email{Matthias.Opel@wmi.badw.de}
 \affiliation{Walther-Mei{\ss}ner-Institut, Bayerische Akademie der
              Wissenschaften, 85748 Garching, Germany}

\author{Rudolf Gross}
\email{Rudolf.Gross@wmi.badw.de}
 \affiliation{Walther-Mei{\ss}ner-Institut, Bayerische Akademie der
              Wissenschaften, 85748 Garching, Germany}
 \affiliation{Physik-Department, Technische Universit\"{a}t M\"{u}nchen, 85748 Garching, Germany}

\date{\today}%

\begin{abstract}
The anomalous Hall effect (AHE) has been studied systematically in the
low-conductivity ferromagnetic oxide Fe$_{3-x}$Zn$_x$O$_4$ with $x = 0$, 0.1,
and 0.5. We used (001), (110), and (111) oriented epitaxial
Fe$_{3-x}$Zn$_x$O$_4$ films grown on MgO and sapphire substrates in different
oxygen partial pressure to analyze the dependence of the AHE on
crystallographic orientation, Zn content, strain state, and oxygen deficiency.
Despite substantial differences in the magnetic properties and magnitudes of
the anomalous Hall conductivity $\sigma_{xy}^{\rm AHE}$ and the longitudinal
conductivity $\sigma_{xx}$ over several orders of magnitude, a universal
scaling relation $\sigma_{xy}^{\rm AHE} \propto \sigma_{xx}^{\alpha}$ with
$\alpha = 1.69 \pm 0.08$ was found for all investigated samples. Our results
are in agreement with recent theoretical and experimental findings for
ferromagnetic metals in the dirty limit, where transport is by metallic
conduction. We find the same scaling relation for magnetite, where hopping
transport prevails. The fact that this relation is independent of
crystallographic orientation, Zn content, strain state, and oxygen deficiency
suggests that it is universal and particularly does not depend on the nature of
the transport mechanism.
\end{abstract}

\pacs{75.47.-m, % Magnetotransport phenomena; materials for magnetotransport
      72.25.-b, % Spin polarized transport
      73.50.-h, % Electronic transport phenomena in thin films
      75.47.Pq} % Magnetotransport phenomena; materials for magnetotransport: Other materials

\maketitle

The physics of the Hall effect in ferromagnetic materials is discussed
intensively and controversially since the 1950s. Early experimental work on
ferromagnetic metals suggested that the Hall resistivity can be described by
the empirical relation $\rho_{xy} = R_{\rm O} \mu_0 H + R_{\rm A} \mu_0 M$, where $H$
is the applied magnetic field and $M$ the spontaneous magnetization of the
ferromagnet. The first term, proportional to $H$ and characterized by the
ordinary Hall coefficient $R_{\rm O}$, describes the ordinary Hall effect (OHE),
whereas the second term, proportional to $M$ and characterized by the anomalous
Hall coefficient $R_{\rm A}$, represents the anomalous Hall effect (AHE). Although
the AHE is generally observed in ferromagnetic metals and semiconductors, its
origin has been one of the most intriguing and controversial issues in
solid-state physics and various theories based on intrinsic and extrinsic
mechanisms have been proposed
\cite{Karplus:1954a,Smit:1955a,Luttinger:1958a,Berger:1970a,Nozieres:1973a}.
Whereas the extrinsic origins of the AHE are based on skew scattering
\cite{Smit:1955a,Luttinger:1958a} and side jump
\cite{Berger:1970a,Nozieres:1973a} mechanisms due to spin orbit interaction
connecting the spin polarization with the orbital motion of electrons, the
intrinsic origin of the AHE is closely related to the Berry phase
\cite{Berry:1984a} of the Bloch electrons
\cite{Karplus:1954a,Sundaram:1999a,Onoda:2002a,Jungwirth:2002a,Burkov:2003a,Nagaosa:2006a,Onoda:2006a,Onoda:2008a}.
The dissipationless and topological nature of the intrinsic mechanism has
attracted much attention recently and various first principles band structure
calculations have been performed to explain the AHE in transition metals
\cite{Yao:2004a,Baily:2005a}, ferromagnetic semiconductors
\cite{Jungwirth:2002a,Burkov:2003a,Zeng:2006a}, and oxides
\cite{Lynada-Geller:2001a,Fang:2003a,Mathieu:2004a,Wang:2006a}.

A powerful experimental test for AHE models is the measurement of the scaling
of the anomalous Hall resistivity (conductivity) $\rho_{xy}^{\rm AHE}$
($\sigma_{xy}^{\rm AHE}$) with the longitudinal resistivity (conductivity)
$\rho_{xx}$ ($\sigma_{xx}$). The skew scattering and side jump mechanisms are
known to yield $\rho_{xy}^{\rm AHE} \propto \rho_{xx}$ ($\sigma_{xy}^{\rm AHE}
\propto \sigma_{xx}$) and $\rho_{xy}^{\rm AHE} \propto \rho_{xx}^2$
($\sigma_{xy}^{\rm AHE} \sim const.$), respectively. Recently, a unified theory
of the AHE has been developed for multiband ferromagnetic metals with dilute
impurities, taking into account resonant contributions from band crossings.
This model predicts three scaling regimes as a function of electron scattering
time. In the clean limit the skew scattering mechanism dominates resulting in
$\sigma_{xy}^{\rm AHE} \propto \sigma_{xx}$. On decreasing scattering time or
conductivity, the intrinsic contribution becomes dominant, yielding
$\sigma_{xy}^{\rm AHE} \simeq const$. In the dirty limit, the intrinsic
contribution is strongly damped, resulting in a scaling relation
$\sigma_{xy}^{\rm AHE} \propto \sigma_{xx}^{1.6}$
\cite{Onoda:2006a,Onoda:2008a}. The crossover between the intrinsic and dirty
limit regime has been observed very recently for several itinerant ferromagnets
\cite{Miyasato:2007a}. Furthermore, the $\sigma_{xy}^{\rm AHE} \propto
\sigma_{xx}^{1.6}$ scaling has been found for several low-conductivity
materials independent of the details of the underlying transport mechanism
(hopping or metallic conduction)
\cite{Miyasato:2007a,Toyosaki:2004a,Ueno:2007a,Fukumura:2007a,Fernandez-Pacheco:2008a}.
This is quite surprising and not understood in detail so far. Although
theoretical models have been proposed for the AHE in low-conductivity
ferromagnetic materials \cite{Burkov:2003a,Lynada-Geller:2001a}, where
transport is dominated by hopping, or in the Hall insulator phase
\cite{Pryadko:1999a}, no complete theoretical understanding has been developed.
In particular, no universal scaling behavior independent of the underlying
transport mechanism has been predicted.

In this Letter, we address the question whether or not the scaling relation
$\sigma_{xy}^{\rm AHE} \propto \sigma_{xx}^{1.6}$ predicted for ferromagnetic
metals in the dirty limit \cite{Onoda:2006a,Onoda:2008a} and quantum Hall
insulators \cite{Pryadko:1999a} also holds for low-conductivity ferromagnetic
oxides with hopping type conductivity as indicated by recent experiments
\cite{Fukumura:2007a,Fernandez-Pacheco:2008a}. To do so we performed a
systematic study of the AHE in magnetite (Fe$_3$O$_4$), which belongs to this
material class. In particular, we checked whether the scaling relation is
universal or depends on the specific sample properties. In order to clarify
possible influences of the crystal structure, the charge carrier density, or Zn
content we studied (i) epitaxial Fe$_3$O$_4$ films with (001), (110), and (111)
orientation, (ii) films grown under different oxygen partial pressure resulting
in different amounts of oxygen vacancies, and (iii) Fe$_{3-x}$Zn$_x$O$_4$ films
with $x=0.1$ and $0.5$. Although the magnetic properties such as saturation
magnetization and coercivity differ significantly among the various samples and
the absolute conductivity values range over four orders of magnitude, the same
scaling relation $\sigma_{xy}^{\rm AHE} \propto \sigma_{xx}^{\alpha}$ with
$\alpha = 1.69\pm 0.08$ is observed for all samples providing evidence that
this scaling relation is indeed universal.

\begin{figure}[b]
    \includegraphics[width=0.9\columnwidth]{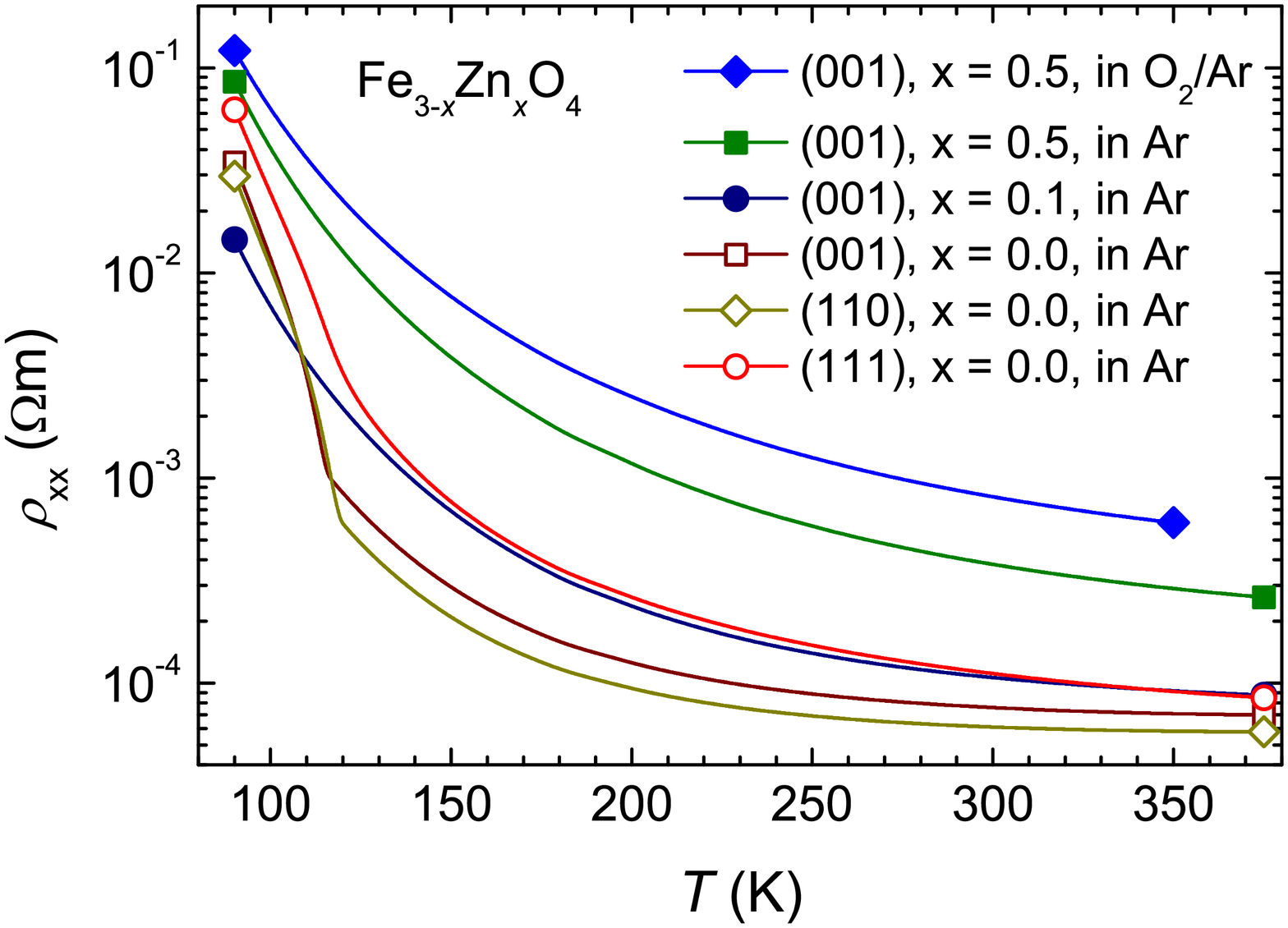}
    \caption{(color online)
             Longitudinal resistivity versus temperature for epitaxial Fe$_{3-x}$Zn$_x$O$_4$ films. The (001), (110), and (111)
             oriented films were grown on MgO(001), MgO(110), and Al$_2$O$_3$(0001) substrates.
             }
    \label{fig:rho}
\end{figure}

Epitaxial thin films of Fe$_{3-x}$Zn$_x$O$_4$ with $x=0$, 0.1, and 0.5 were
grown by laser molecular beam epitaxy \cite{Gross:2000a,Klein:1999b} at a base
pressure of $3.7 \times 10^{-3}$\,mbar in pure Ar atmosphere or an Ar/O$_2$
(99:1) mixture. We used MgO(001), MgO(110), and Al$_2$O$_3$(0001) substrates to
obtain (001), (110), and (111) oriented films with different amount of
epitaxial coherency strain, respectively. Details of the growth process are
given elsewhere\cite{Reisinger:2003a,Reisinger:2003b,Reisinger:2004a}. X-ray
diffraction reveals a high epitaxial quality of our samples as demonstrated by
a FWHM of the rocking curves of the (004) reflection smaller than 0.05$^\circ$.
X-ray reflectometry was used to get precise values of the film thickness
ranging between 30 and 60\,nm. Magnetic characterization was performed by SQUID
magnetometry. For magnetotransport measurements the films were patterned into
typically $45\,\mu$m wide and $350\,\mu$m long Hall bars by optical lithography
and Ar ion beam milling \cite{Alff:1992a}. The Hall bars were aligned parallel
to the in-plane $\langle 100 \rangle$ directions for the (001) and (110) films
and also along the $\langle 110 \rangle$ directions for the (110) films, and
parallel to the [1-10] and [11-2] directions for the (111) films.

Magnetotransport measurements as a function of temperature $T$ and magnetic
field $H$ were performed using a standard 4-probe technique with $H$ applied
perpendicular to the film plane. The Hall resistivity $\rho_{xy}$ was obtained
by anti-symmetrization of the values measured for opposite magnetic field
directions to eliminate offsets due to contact potential or a geometric
misalignment of the Hall probes. The respective conductivities were derived by
inversion of the resistivity tensor. For cubic symmetry and $H \| z$, we get
\begin{eqnarray}
\sigma_{xx} & = & \frac{\rho_{xx}}{\rho_{xx}^2 + \rho_{xy}^2} \simeq 1/\rho_{xx}
 \label{sigmaxx} \\
\sigma_{xy} & = & \frac{\rho_{xy}}{\rho_{xx}^2 + \rho_{xy}^2}  \simeq \frac{\rho_{xy}}{\rho_{xx}^2} \;\; ,
 \label{sigmaxy}
\end{eqnarray}
since $\rho_{xy} \ll \rho_{xx}$ for our samples.

We first discuss $\rho_{xx}(T)$ for the different films. As shown in
Fig.~\ref{fig:rho}, $\rho_{xx}$ increases by more than two orders of magnitude
on decreasing $T$ from 350 to below 100\,K. Furthermore, $\rho_{xx}$
sensitively depends on the Zn-doping, the growth atmosphere and the
crystallographic orientation. An increasing Zn content increases resistivity.
This is caused by the substitution of Fe$^{3+}$ by Zn$^{2+}$ ions reducing the
carrier concentration \cite{Takaobushi:2006a,Takaobushi:2007a}. Zn-substitution
also smears out the Verwey transition \cite{Verwey:1939a} which is clearly seen
for the undoped samples as a pronounced increase of resistivity at $T_V \simeq
(115\pm 5)$\,K. This temperature is slightly smaller than the ideal bulk value
(120\,K) and also the increase in $\rho_{xx}$ is less sharp than for bulk
samples what can be attributed to the nonvanishing epitaxial strain
\cite{Gupta:1998a}. It is also seen that the resistivity increases by growing
the films in an Ar/O$_2$ mixture. This is due to a reduction of oxygen
vacancies resulting in a decrease of the carrier density.

\begin{figure}[tb]
    \includegraphics[width=0.95\columnwidth]{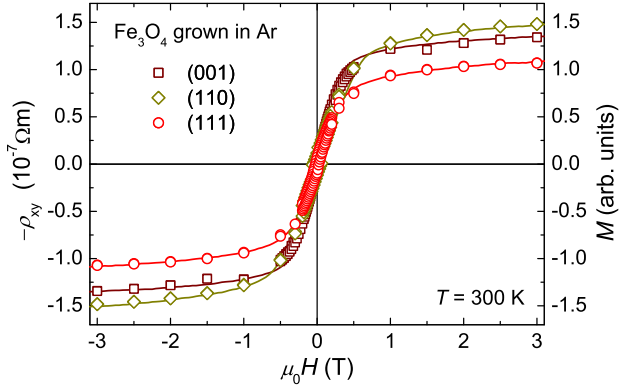}
    \caption{(color online)
             Room temperature Hall resistivity (lines) and magnetization (symbols) plotted versus the magnetic field applied
             perpendicular to the film plane for epitaxial (001), (110), and (111)
             oriented Fe$_{3}$O$_4$ films grown on MgO(001), MgO(110), and Al$_2$O$_3$(0001) substrates.
             }
    \label{fig:AHE}
\end{figure}

\begin{figure}[b]
    \includegraphics[width=0.9\columnwidth]{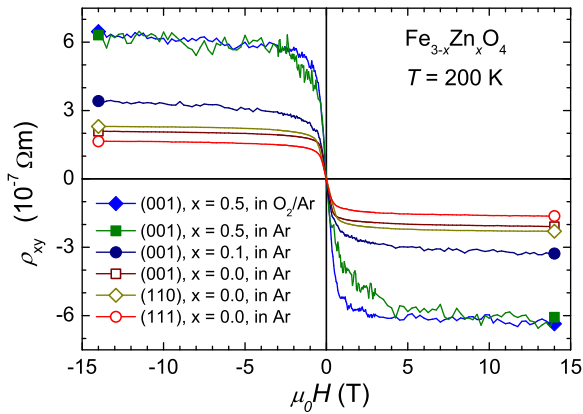}
    \caption{(color online)
             Hall resistivity $\rho_{xy}$ plotted versus magnetic field applied perpendicular
             to the film plane at 200\,K for six different Fe$_{3-x}$Zn$_x$O$_4$ films.}
    \label{fig:RH}
\end{figure}

The Hall resistivity $\rho_{xy}$ is plotted versus $H$ in Fig.~\ref{fig:AHE} at
$T=300$\,K together with the magnetization curves of the same films. Similar
curves are obtained at other temperatures. It is evident that $\rho_{xy}$
scales with magnetization $M$. This is expected according to the empirical
relation $\rho_{xy} = \rho_{xy}^{\rm OHE} + \rho_{xy}^{\rm AHE} = R_{\rm O} \mu_0 H +
R_{\rm A} \mu_0 M$, if the anomalous contribution $\rho_{xy}^{\rm AHE}$ dominates. In
Fig.~\ref{fig:RH} we have plotted $\rho_{xy}(H)$ at 200\,K up to 14\,T. The
data show the already discussed behavior following the $M(H)$ curves due to the
dominating anomalous Hall contribution. We also note that the initial slope of
the $\rho_{xy}(H)$ curves can be larger than $1\,\mu\Omega$m/T corresponding to
a sensitivity of more than 100\,V/AT for a 10\,nm thick film in agreement with
a recent report \cite{Fernandez-Pacheco:2008a}. To derive $\rho_{xy}^{\rm
AHE}(M_{\rm s})= R_{\rm A} \mu_0 M_{\rm s}$, where $M_{\rm s}$ is the saturation magnetization, from
the measured $\rho_{xy}(H)$ curve, one has to separate the ordinary and
anomalous contributions. If $M$ saturates at high fields, this can be simply
done by extrapolation of the linear high field part of $\rho_{xy}(H)$ back to
$H=0$, yielding $\rho_{xy}^{\rm AHE}(H=0) = \rho_{xy}^{\rm AHE}(M=M_{\rm s})$.
Unfortunately, such straightforward analysis is not possible for magnetite,
since $M$ does not fully saturate up to fields of the order of 10\,T due to
anti-phase boundaries \cite{Margulies:1997a}. Fortunately, $\rho_{xy}^{\rm OHE}
\sim 2\times 10^{-9}\,\Omega$m at 1\,T
\cite{Fernandez-Pacheco:2008a,Reisinger:2004a} is very small and therefore can
be safely neglected compared to the about two orders of magnitude larger
contribution $\rho_{xy}^{\rm AHE}$ of the AHE. Then, in good approximation we
can use $\rho_{xy}^{\rm AHE}(H=0)=\rho_{xy}^{\rm AHE}(M_{\rm s}) \simeq
\rho_{xy}(14\,\textrm{T})$. The corresponding anomalous Hall conductivity is
derived as $\sigma_{xy}^{\rm AHE}=\rho_{xy}^{\rm AHE}(0)/\rho_{xx}^2(0)$.

\begin{figure}[tb]
    \includegraphics[width=0.9\columnwidth]{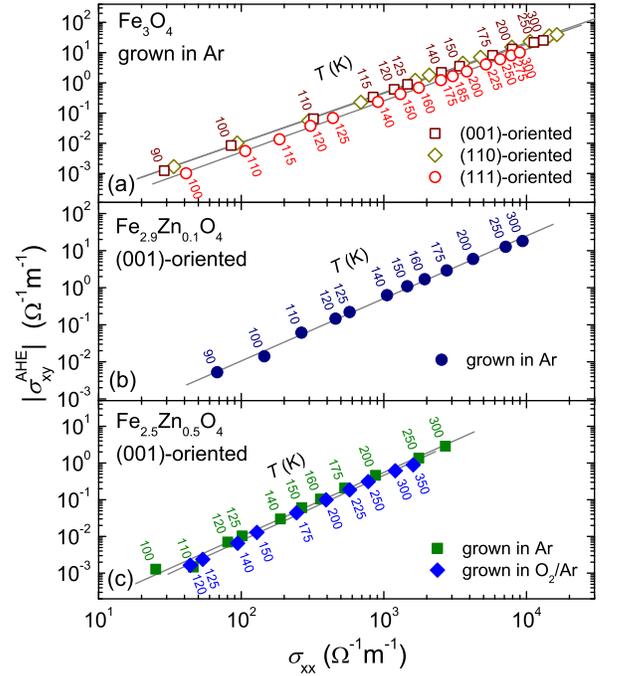}
    \caption{(color online)
             Modulus of the anomalous Hall conductivity, $|\sigma_{xy}^{\rm AHE}|$, plotted versus longitudinal
             conductivity $\sigma_{xx}$ in a double logarithmic representation for different epitaxial
             Fe$_{3-x}$Zn$_x$O$_4$ films in the temperature regime between 90 and 350\,K.
             The lines are linear fits to the data.}
    \label{fig:scaling}
\end{figure}

By varying the crystallographic orientation, the growth atmosphere, and the Zn
content we have fabricated epitaxial Fe$_{3-x}$Zn$_x$O$_4$ films with
$\sigma_{xy}^{\rm AHE}$ and $\sigma_{xx}$ values ranging over almost five and
three orders of magnitude, respectively, in the studied temperature regime from
90 to 350\,K. The most intriguing result is that irrespective of these
pronounced differences all data follow the same scaling plot (see
Fig.~\ref{fig:scaling})
\begin{equation}
\left| \sigma_{xy}^{\rm AHE} \right|  =  a \cdot \sigma_{xx}^\alpha
 \label{scaling}
\end{equation}
with $\alpha = 1.69 \pm 0.08$ and $a=(4 \pm 2) \times 10^{-6} [1/\Omega
\textrm{m}]^{1-\alpha}$. This scaling behavior even holds on moving across the
Verwey transition, which is associated with a structural phase transition
(cubic to monoclinic) as well as charge and/or orbital ordering
\cite{Garcia:2001a,Walz:2002a}. With respect to the scaling exponent $\alpha$
our systematic data is in good agreement with theoretical predictions
\cite{Onoda:2006a,Onoda:2008a}. Some data even indicate a tendency towards
saturation of $\sigma_{xy}^{\rm AHE}$ at conductivity values above
$10^4\Omega^{-1}\textrm{m}^{-1}$. Our results also agree well with those
obtained on polycrystalline \cite{Feng:1975a} and epitaxial Fe$_3$O$_4$ films
of different thickness \cite{Fernandez-Pacheco:2008a} as well as those obtained
from single crystals \cite{Todo:1995a}. We further note that scaling relations
with about the same exponent but different prefactors have been found for
Ti$_{1-x}$Co$_x$O$_{2-\delta}$ \cite{Toyosaki:2004a} and a number of other
oxide materials
\cite{Miyasato:2007a,Fukumura:2007a,Lynada-Geller:2001a,Ueno:2007a}. This
strongly suggests that the observed scaling behavior in low-conductivity
ferromagnetic oxides is universal and not related to details of the crystal
structure and crystallographic orientation, doping level, oxygen stoichiometry,
strain state, or density of anti-phase boundaries.

The origin of the observed universal scaling relation is still controversial.
Although a recent theoretical model \cite{Onoda:2006a,Onoda:2008a} developed
for multiband ferromagnetic metals with dilute impurities predicts
$\sigma_{xy}^{\rm AHE} \propto \sigma_{xx}^{1.6}$ in the dirty limit, this
model does not cover the case of low-conductivity ferromagnetic oxides where
electric transport is dominated by hopping. Furthermore, it is still unclear to
what extent the model developed for quantum Hall insulators
\cite{Pryadko:1999a} can be applied. In any case, our results together with the
available literature data
\cite{Miyasato:2007a,Toyosaki:2004a,Fukumura:2007a,Fernandez-Pacheco:2008a}
suggest that the scaling relation holds for low-conductivity materials
independent of the details of the specific material parameters and the
underlying transport mechanism (hopping or metallic conduction) and therefore
can be considered universal.

In summary, we have performed a systematic study of the AHE in the
low-conductivity ferromagnetic oxide Fe$_{3-x}$Zn$_x$O$_4$.  Independent of the
crystallographic orientation, Zn content, oxygen deficiency, and strain state,
resulting in substantial differences in magnetic and magnetotransport
properties, a universal scaling relation $\sigma_{xy}^{\rm AHE} \propto
\sigma_{xx}^{\alpha}$ with $\alpha = 1.69 \pm 0.08$ was found. Comparing our
results with recent literature data we can conclude that the observed scaling
relation is very general and holds for low-conductivity ferromagnetic materials
irrespective of the details of the underlying transport mechanism.

Financial support by the Deutsche Forschungsgemeinschaft via the priority
programs 1157 and 1285 (project Nos. GR~1152/13+14) and the Excellence Cluster
"Nanosystems Initiative Munich (NIM)" is gratefully acknowledged. D.V. and
M.S.R.R. thank the DAAD for financial support. We also acknowledge technical
support by A. Erb.

\end{document}